# Thermoelectric characterization of fine-grained Ti$_5$O$_9$ Magnéli phase ceramics

S. Pandey,[1] G. Joshi,[2] S. Wang,[3] S. Curtarolo,[3] and R. Gaume[4, 5, 6]

[1]*Department of Physics, University of Central Florida, Orlando, Florida, 32816, USA*

[2]*Evident Thermoelectrics, Troy, New York, 12180, USA*

[3]*Department of Material Science and Engineering, Duke University, Durham, North Carolina, 27708, USA*

[4]*College of Optics and Photonics, CREOL, University of Central Florida, Orlando, Florida, 32816, USA*

[5]*Department of Material Science and Engineering, University of Central Florida, Orlando, Florida, 32816, USA*

[6]*NanoScience Technology Center, University of Central Florida, Orlando, Florida, 32816, USA*

Magnéli phase Ti$_5$O$_9$ ceramics with 200-nm grain-size were fabricated by hot-pressing nanopowders of titanium and anatase TiO$_2$ at 1223 K. The thermoelectric properties of these ceramics were investigated from room temperature to 1076 K. We show that the experimental variation of the electrical conductivity with temperature follows a small-polaron model and that the Seebeck coefficient can be explained by a temperature-dependent Heikes-Chaikin-Beni model. The thermoelectric figure-of-merit ZT of this nanoceramic material reaches 0.3 at 1076 K.

Thermoelectric (TE) devices convert heat into electricity and are used in waste heat recovery and power generation. Their performance is evaluated using the dimensionless figure-of-merit ZT defined by $ZT = \frac{\sigma \alpha^2}{\kappa}T$, where σ, α, κ and T are the electrical conductivity, thermopower (Seebeck coefficient), thermal conductivity and temperature respectively. With ZT values between 1 and 2.6[1-3], state-of-the-art TE materials are generally based on chalcogenide and antimonide materials. On the other hand, and despite their lower ZT values, oxide TE materials offer less toxic, cheaper and more chemically stable alternatives. In particular, their high temperature stability allows oxide TE device to be operated at larger temperatures and subjected to higher temperature gradients thereby increasing their efficiency and somewhat compensating for their lower intrinsic ZT figures[3].

Magnéli phase titanium oxides, with generic formula Ti$_n$O$_{2n-1}$, possess some desirable qualities for TE applications. Their crystal structure derives from that of rutile and display oxygen vacancies in every n$^{th}$ titanium octahedral site[4]. These vacancies, combined with large unit cells, increase phonon scattering and hence reduce the lattice contribution towards thermal conductivity. In nanoceramics, the thermal conductivity is further decreased by achieving nanometer-scale grain-sizes that enhance phonon scattering at the grain boundaries. On the other hand, the large concentration of $3d^1$ $Ti^{3+}$ contributes to enhancing the electrical conduction and thermopower. Slight stoichiometry variations in $Ti_nO_{2n-1}$ have large effects on the material band structure[5] and, with suitable doping, the thermoelectric properties can be tuned easily. The TE properties of titanium suboxides, including that of $Ti_5O_9$, have been previously investigated and ZT values up to 0.4 have been obtained at 1150 K[4,6-12]. These phases however have not been extensively characterized and modelled. Here, we report on the analysis of experimental TE measurements with polaron models and band-structure calculations.

Ceramic samples of $Ti_5O_9$ were prepared by solid-state reaction in a hot-press between nanopowders of anatase $TiO_2$ and titanium. 50 nm anatase $TiO_2$ powders (Sigma-Aldrich) and 30-nm Ti powders (American Elements) were ball-milled together in a 39:5 mass ratio in the presence of acetone for 20 hours. The mixture was then dried, sieved and sintered in a hot-press at 950°C and 40 MPa, under argon in a graphite die assembly. The heating and cooling rates were 8°C/min and 15°C/min, respectively. The holding time was varied between 0, 0.5, 2, 6 and 10 hrs. The phase composition and microstructure of the ceramic discs were characterized by x-ray diffraction (Rigaku D/Max, Cu-Kα radiation) and scanning electron microscopy (Zeiss Ultra-55 SEM).

The thermal conductivity was measured on polished sintered samples with a laser-flash instrument (Netzsch, LFA467), while the Seebeck coefficient and electrical conductivity were measured on a commercially available equipment (ULVAC, ZEM-3).

X-ray diffraction reveals that the formation of the $Ti_5O_9$ Magnéli phase is complete at 950°C and that minor amounts of Magnéli phase $Ti_4O_7$ is present (Fig. 1). The poor counting rate on this measurement may be due to the low structure factor associated to the large $Ti_5O_9$ unit cell and to the fact that the long-range ordering of oxygen vacancies, characteristic of Magnéli phases, is not yet complete. This interpretation is corroborated by the fact that, under the same acquisition conditions, higher peak intensities are obtained for longer sintering times (Fig. 1).

The mixing ratio of the reactants corresponds to nearly twice as much titanium as would be stoichiometrically required to form $Ti_5O_9$. This recipe, documented in prior work and redeveloped through our own optimizations, suggests that the titanium nanopowders are coated by a thin shell of $TiO_2$, shifting the stoichiometry of the reaction towards the higher oxidation Magnéli phases. Based on this assumption and on the mixing ratio utilized for the synthesis, we expect a 2.4-nm thick oxide layer on our 30-nm titanium particles in order to produce $Ti_5O_9$. This estimate is consistent with some direct observation of passivation layers on Ti thin-films[13] and with our own transmission electron microscopy (TEM) and electron energy loss spectroscopy (EELS) analyses which reveal a 2- to 4-nm thick oxide layer around the titanium nanoparticles (Fig. 2).

After sintering, the finest microstructure obtained in our study consisted of 160-nm grains with an estimated porosity of 3.3 %, mostly composed of 105-nm size pores. It was obtained by cooling the hot-press immediately after reaching the sintering temperature. As expected, the microstructure slowly coarsens with increasing dwell times (Fig. 3).

The thermoelectric characterizations that follow were performed on a sample hot-pressed for 30 min, since these conditions yield both a dense and a fine microstructure.

Figure 4 shows the temperature dependence of the electrical conductivity. Like in most transition metal oxides, electrical conduction in Magnéli phases occurs by polaron hopping[14-16] and electron transfer between the $t_{2g}$ levels of two neighbouring titanium ions is mediated by bridging oxygen ions. Thus, the average hopping distance, R, corresponds to the length of the Ti-O-Ti bonds and is equal to 3.8 Å in $Ti_5O_9$[5]. In comparison, this titanium suboxide, with a polaron density of $N=1.25 \cdot 10^{28}$ m$^{-3}$, has an estimated polaron size of[17]:

$$2\,r_p = (\frac{\pi}{6N})^{1/3} \sim 3.46\ \text{Å} \qquad (1)$$

This shows that the temperature dependency of the electronic conduction in $Ti_5O_9$ can be approached by small-polaron models such as Holstein's (Eq. 2), if the energy difference, $W_D$, between two neighbouring localized states is zero[17], or Schnakenberg's (Eq. 3), if $W_D \neq 0$ [17,18]:

$$\sigma = \frac{\sigma_0}{T^m} e^{\left(-\frac{W}{K_B T}\right)} \qquad (2)$$

$$\sigma = \frac{\sigma_0}{T}\sqrt{\sinh(\frac{\hbar\omega_0}{k_B T})}\, e^{\left(-\frac{4W_H}{\hbar\omega_0}\tanh(\frac{\hbar\omega_0}{4k_B T})\right)} e^{\left(-\frac{W_D}{k_B T}\right)} \qquad (3)$$

In these expressions, $\sigma_0$ is a constant, $W_H$ is the polaron hopping energy, $\omega_0$ the longitudinal optical phonon frequency and $W=W_H+W_D/2$ the activation energy for hopping[17]. The adiabaticity exponent m of Holstein's model characterizes the electron transfer probability between two hopping sites. A high transfer probability (adiabatic case) corresponds to m=1, whereas a low transfer probability (non-adiabatic case) has m=1.5. Despite their similar environments, the titanium sites in $Ti_5O_9$ are not all crystallographically equivalent and it is expected that, while small in magnitude, $W_D$ might be non-zero. For this reason, we have fitted our experimental data to both models (Fig. 4).

The fit to Eq. 2 yields W=37 and 61 meV for the adiabatic and non-adiabatic cases respectively. These values are comparable to the activation energies found in other transition metal oxides[19-21].

On the other hand, since R is not much larger than $r_p$, the value of $W_H$ in the Schakensberg model can be estimated by[22]:

$$W_H = \frac{e^2}{4}(\frac{1}{\epsilon_\infty} - \frac{1}{\epsilon_s})(\frac{1}{r_p} - \frac{1}{R}) \qquad (4)$$

where $\epsilon_\infty$ and $\epsilon_s$ are the high frequency and static permittivities of the material and e is the charge of an electron. Taking the $\epsilon_s$ and $\epsilon_\infty$ values of $TiO_2$[15], $W_H$ is found to be on the order of 60 meV and the fit to the conductivity data using Eq. 3 yields $\omega_0 = 10^{13}$ Hz, a typical value for oxide materials[23]. As anticipated, we also find small $W_D$ value of 7 meV from which we deduce an activation energy for polaron hopping of W=64 meV, a value in agreement with the fit using Holstein's model with m=1.5.

The non-adiabatic character of the electron transfer can further be tested against Holstein's criterion[14,23] by verifying that the polaron energy bandwidth, J, is smaller than the quantity:

$$\varphi = \left(\frac{2k_B T W_H}{\pi}\right)^{\frac{1}{4}} \left(\frac{\hbar\omega_0}{\pi}\right)^{\frac{1}{2}} = 21 \text{ meV} \qquad (5)$$

We have used *ab-initio* calculations to determine the density of states of $Ti_5O_9$ at the Fermi level, $N(E_F)$, (Fig. 5) and derived the polaron energy bandwidth, J, from Mott and Davis formula[17]:

$$J \sim e^3 N(E_F)^{\frac{1}{2}} \left(\frac{1}{\epsilon_\infty} - \frac{1}{\epsilon_s}\right)^{3/2} \qquad (5)$$

With $N(E_F)=5\cdot10^{23}$ eV/m³, we find $J\sim3$ meV$<\varphi$ , a value which ascertains the non-adiabatic behavior of polaron hopping in $Ti_5O_9$.

The Seebeck coefficient is defined as the entropy per charge carrier and is usually written as[24]:

$$\alpha = \alpha_{presence} + \alpha_{transport} \qquad (6)$$

The first term corresponds to on-site disorder (including configuration, spin degeneracy and vibrational effects), while the second term accounts for charge carrier transport through the lattice. Neglecting the vibrational contribution, the first term can be approximated by the Heikes-Chaikin-Beni (HCB) formula[25] and written as a function of the charge carrier density n:

$$\alpha_{presence} = \frac{k_B}{e} \ln\left[\frac{g_{Ti^{4+}}}{g_{Ti^{3+}}}\left(\frac{n}{1-n}\right)\right] \qquad (7a)$$

or, in the case of undoped $TiO_x$, as a function of the stoichiometric coefficient x:

$$\alpha_{presence} = \frac{k_B}{e} \ln\left[\frac{g_{Ti^{4+}}}{g_{Ti^{3+}}}\left(\frac{4-2x}{2x-3}\right)\right] \qquad (7b)$$

In these expressions, $g_{Ti^{3+}}$ and $g_{Ti^{4+}}$ are the spin-orbital degeneracies of the lowest-lying state of $Ti^{3+}$ and $Ti^{4+}$ species. In $Ti_5O_9$, tetragonal distortion of the $Ti^{3+}$ sites produces a compressed octahedral crystal-field environment[26,27] that lifts the degeneracy of the $e_g$ states into two singlets ($b_1$ and $a_1$), lying in the conduction band, and of the $t_{2g}$ states into one doublet (e=$d_{xz}$, $d_{yz}$) and one singlet ($b_2$=$d_{xy}$), both residing in the band gap (Fig. 5 and 7). With one single electron on the $b_2$ state, one finds $g_{Ti^{3+}}=2$, $g_{Ti^{4+}}=1$ and the Seebeck coefficient of undoped $Ti_5O_9$ (x=1.8 i.e. n=4-2x=0.4) is expected to be -94.5 µV/K. Figure (6) shows the graph of Eq. (7) along with our experimental measurement and data from the literature, all taken near ambient temperature[6-8,10]. While the general trend of the HCB formula agrees with the experimental data, quantitative agreement is only approximate most likely because of the nature of the approximations made in the model, as well as possible contributions from donor/acceptor impurities and experimental variance in the determination of x and α by the various investigators.

The temperature dependency of the thermopower is often discussed in the context of vibrational entropy transport[24]. For example, in the case of small polaron hopping, the second part of Eq. (6) is given by[24]:

$$\alpha_{transport} = \frac{1}{eT} \left\langle \frac{zJ^2 k_B}{2E_B^3} \left(E_f - E_i\right)^2 \right\rangle \qquad (8)$$

where $E_f$ and $E_i$ are the energies of the system when the hopping sites i and f are occupied, z the coordination number around the polaron site and $E_B$ the polaron binding energy. However, despite the presence of inequivalent crystallographic titanium sites, this transport term is expected to be small in $Ti_5O_9$ due to the small value of J found previously and amount to about $\alpha_{transport}\sim 2.6\cdot 10^{-2}$ µV/K at 300 K and $7.2\cdot 10^{-2}$ µV/K at 800 K. In contrast, we propose that, in the particular case of $Ti_5O_9$, the relative proximity of the doubly-degenerate e state from the ground-state $b_2$ gives rise to the possibility of increasing the carrier entropy through direct thermal excitations in the temperature range investigated here. This effect can be modelled analytically using the partition function, Z, of electrons at equilibrium between the three $t_{2g}$ states of Ti species, neglecting the high-energy $a_1$ and $b_1$ states:

$$Z = 1 + 2g_{b_2}e^{-\beta\mu} + 2g_e e^{-\beta(\mu+\Delta)} \quad (9)$$

In this expression, the first term corresponds to the contribution of $Ti^{4+}$, β stands for 1/kT, µ is the electronic potential (Fermi energy) and Δ the energy separation between the $b_2$ and e states of respective orbital degeneracies $g_{b2}=1$ and $g_e=2$. The factors of two in this equation account for the spin multiplicity for these singly-occupied states. Hence, the average carrier density, n, is:

$$n = \frac{2g_{b_2}e^{-\beta\mu} + 2g_e e^{-\beta(\mu+\Delta)}}{Z} \quad (10)$$

and Eq. (10) can be solved for µ:

$$\mu = -kT\ln\left[\frac{n}{2(1-n)(g_{b_1}+g_e e^{-\beta\Delta})}\right] \quad (11)$$

Finally, the Seebeck coefficient is obtained from Eq. (11) using:

$$\alpha = -\frac{1}{e}\left(\frac{\partial\mu}{\partial T}\right)_n \quad (12)$$

$$\alpha = \frac{k}{e}\ln\left[\frac{n}{2(1-n)}\right] - \frac{k}{e}\left[\ln(g_{b_1} + g_e e^{-\beta\Delta}) + \frac{\Delta}{kT}\frac{g_e e^{-\beta\Delta}}{g_{b_1}+g_e e^{-\beta\Delta}}\right] \quad (13)$$

This last relation, in which the temperature dependency is contained in the second term, is analogous to Eq. (7a) in that it extends the HCB formula for the configuration entropy of a canonical ensemble made of closely separated energy levels (Δ/kT ~ 1). When fitted to our experimental data, and with no other constraints than the values of $g_{b2}=1$ and $g_e=2$, we find a remarkable agreement for n=0.516 (x=1.742) and Δ=151 meV. While the value of Δ is consistent with the energy splitting of the $t_{2g}$ levels modeled by Curtarolo et al. (Fig. 5), the fitted value of n exceeds that of undoped $Ti_5O_9$. This result indicates that our material might be lightly doped with donor impurities or, most probably, underwent some slight reduction and evolved into a higher $Ti_4O_7$ (x=1.75) content.

The thermal conductivity of $Ti_5O_9$ ceramics increases with temperature (Fig. 8). The thermal conductivity is the sum of contributions from electrons and phonons, $\kappa = \kappa_e + \kappa_p$. According to the Wiedemann-Franz law, the electronic contribution to thermal conductivity is $\kappa_e = L\sigma T$, where L is the Lorenz number, a material's constant of the order of[28] $10^{-8}$. The value of $\kappa_p$, on the other hand, is proportional to $cvl$, where c is the heat capacity of the solid, v the mean velocity of phonons (which is approximately equal to the sound velocity) and $l$ is the mean free path of phonons. Above the Debye temperature $\theta_D$, c assumes a constant value and $l \propto T^{-1}$ [29]. A trend of the form $\kappa = L\sigma T + A/T$ is expected for the dependency of the thermal conductivity with temperature. Our data suggest that the phonon contribution remains constant in the range of temperatures between 300 and 800°C, and that the thermal conductivity has the form: $\kappa = L\sigma T + A$.

The fit to the data yields a Lorenz number of L=$5.6 \cdot 10^{-9}$ W·Ω·K$^{-2}$ and $\kappa_p = A = $ 2.1 W·m$^{-1}$·K$^{-1}$. This implies that the phonon mean free-path is independent of temperature. This behavior, analogous to that of amorphous materials, is likely the consequence of poor oxygen vacancy ordering (as revealed by x-ray diffraction) and the relatively fine microstructure of the ceramic sample.

The TE figure-of-merit, ZT, can be obtained from the previous data. Thermal conductivity values were extrapolated for 876 K, 975 K and 1076 K to match the range of thermopower and electrical conductivity data. Fig. 9 shows the evolution of ZT as a function of temperature and a value of 0.3 is expected at 1076 K. We expect that larger values of ZT could be achieved in optimized versions of this material with suitable doping to increase charge carrier density.

In conclusion, Magnéli phase $Ti_5O_9$ ceramics were prepared by solid-state reaction in a hot-press between nanopowders of anatase $TiO_2$ and titanium at 950°C and 40 MPa. The Seebeck coefficient and electrical conductivity were measured from 307 K to 1076 K and the thermal conductivity was measured from room temperature to 776 K. The variation of electrical conductivity with temperature is consistent with a small polaron hopping mechanism. Our temperature-dependent Seebeck measurements suggest that the vibrational entropy transported during polaron hopping is negligible and that our data can be explained using a modified Heikes-Chaikin-Beni formula taking into account the configuration entropy of a canonical ensemble made of closely separated $t_{2g}$ levels. The thermal conductivity increases with temperature, and a Lorenz number of $5.6 \cdot 10^{-9}$ W·Ω·K$^{-2}$ was deduced from our data. Such properties yield a ZT value of 0.3 at 1076 K. It is likely that better performance could be obtained in this material system by optimizing doping and by further reducing grain size using techniques such as spark-plasma sintering (SPS).

**References**

[1] X. F. Zheng, C. X. Liu, Y. Y. Yan, and Q. Wang, Renewable and Sustainable Energy Reviews **32,** 486 (2014).

[2] L.-D. Zhao, S.-H. Lo, Y. Zhang, H. Sun, G. Tan, C. Uher, C. Wolverton, V. P. Dravid, and M. G. Kanatzidis, Nature (London) **508,** 373 (2014).

[3] J. He, Y. F. Liu, and R. Funahashi, Journal of Materials Research **26,** 1762 (2011).

[4] S. Harada, K. Tanaka, and H. Inui, Journal of Applied Physics **108** (2010).

[5] S. Curtarolo, in *Automatic-FLOW for materials discovery.*

[6] Q. He, Q. Hao, G. Chen, B. Poudel, X. Wang, D. Wang, and Z. Ren, Applied Physics Letters **91,** 52505 (2007).

[7] Y. Lu, M. Hirohashi, and K. Sato, Materials Transactions **47,** 1449 (2006).

[8] Y. Lu, Y. Matsuda, K. Sagara, L. Hao, T. Otomitsu, and H. Yoshida, Advanced Materials Research **415-417,** 1291 (2011).

[9] Y. Lu, K. Sagara, L. Hao, Z. Ji, and H. Yoshida, MATERIALS TRANSACTIONS **53,** 1208 (2012).

[10] M. Backhaus-Ricoult, J. R. Rustad, D. Vargheese, I. Dutta, and K. Work, Journal of Electronic Materials**,** 1636 (2012).

[11] M. Mikami and K. Ozaki, in *Thermoelectric properties of nitrogen-doped TiO$_{2-x}$ compounds*, 2012 (IOP Publishing), p. 012006.

[12] D. Portehault, V. Maneeratana, C. Candolfi, N. Oeschler, I. Veremchuk, Y. Grin, C. Sanchez, and M. Antonietti, ACS Nano **5,** 9052 (2011).

13 S. Roessler, R. Zimmermann, D. Scharnweber, C. Werner, and H. Worch, Colloids and Surfaces B: Biointerfaces **26,** 387 (2002).

14 A. J. Bosman and H. J. van Daal, Advances in Physics **19,** 1 (1970).

15 N. Tsuda, *Electronic conduction in oxides*, Vol. 94 (Springer, 2000).

16 A. Banerjee, S. Pal, E. Rozenberg, and B. K. Chaudhuri, Journal of Physics: Condensed Matter **13,** 9489 (2001).

17 N. F. Mott and E. A. Davis, *Electronic processes in non-crystalline materials* (OUP Oxford, 2012).

18 J. Schnakenberg, physica status solidi (b) **28,** 623 (1968).

19 S. Li, R. Funahashi, I. Matsubara, K. Ueno, S. Sodeoka, and H. Yamada, Chemistry of materials **12,** 2424 (2000).

20 J. Lago, P. Battle, M. Rosseinsky, A. Coldea, and J. Singleton, Journal of Physics: Condensed Matter **15,** 6817 (2003).

21 Y. Sun, X. Xu, and Y. Zhang, Journal of Physics: Condensed Matter **12,** 10475 (2000).

22 I. Austin and N. F. Mott, Advances in Physics **18,** 41 (1969).

23 A. Yildiz, S. B. Lisesivdin, M. Kasap, and D. Mardare, Physica B: Physics of Condensed Matter **404,** 1423 (2009).

24 C. W. a. D. Emin, Physical Review B **29** (1984).

25 P. Chaikin and G. Beni, Physical Review B **13,** 647 (1976).

26 E. Stoyanov, F. Langenhorst, and G. Steinle-Neumann, American Mineralogist **92,** 577 (2007).

27 V. M. Khomenko, K. Langer, H. Rager, and A. Fett, PHYSICS AND CHEMISTRY OF MINERALS **25,** 338 (1998).

28 G. S. Kumar, G. Prasad, and R. O. Pohl, Journal of Material Science **28,** 4261 (1993).

29 C. Wood, Reports on Progress in Physics **51,** 459 (1988).

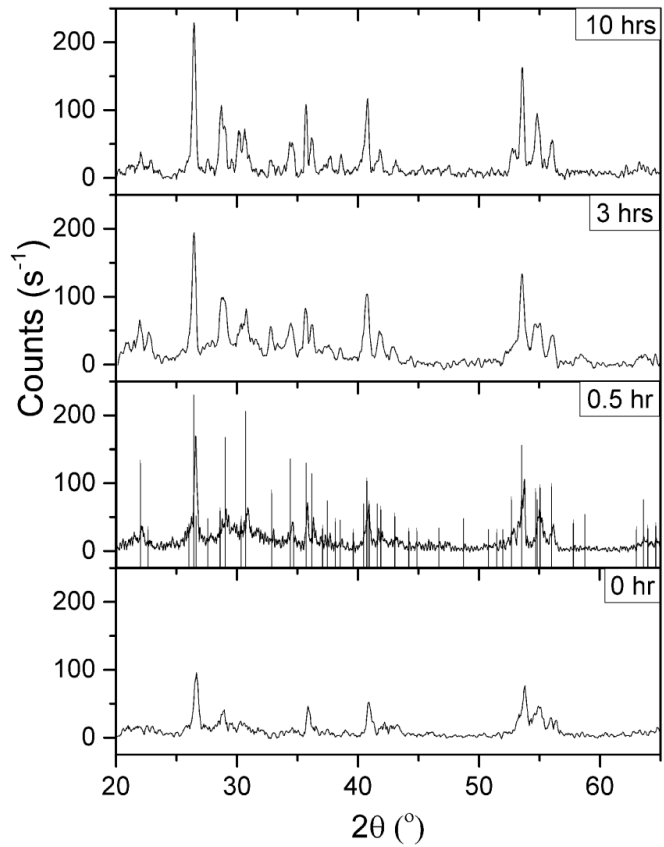

**Figure 1** X-ray diffractogram of a sample hot-pressed at 950°C for 0, 0.5, 3 and 10 hours. The vertical lines correspond to the reference pattern of $Ti_5O_9$ (PDF#051-0641).

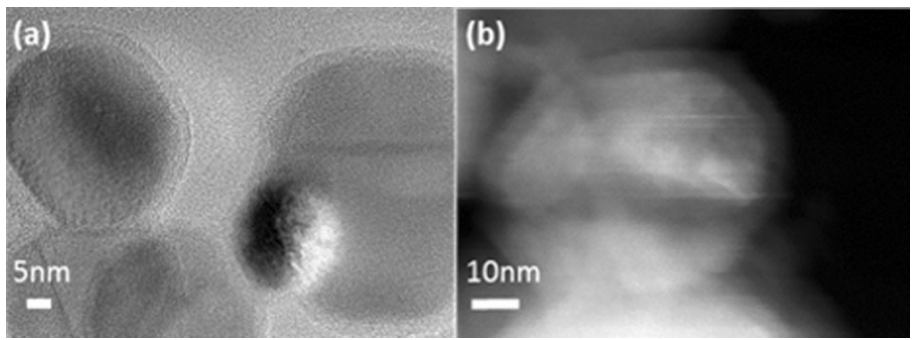

**Figure 2** TEM images of titanium nanoparticles. Figure (b) is taken slightly out of focus to obtain better contrast between the titanium core and the titanium oxide shell.

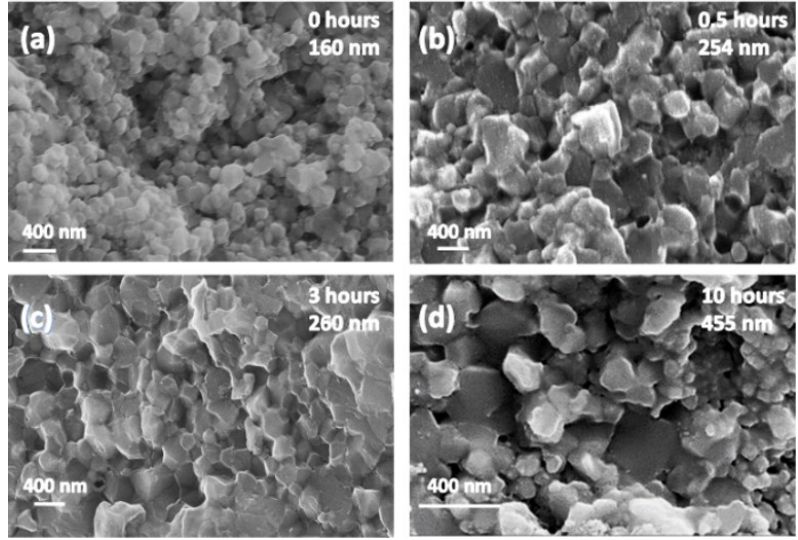

**Figure 3** Microstructures of ceramic samples hot-pressed at 950°C for (a) 0, (b) 0.5, (c) 3 and (d) 10 hours. The average grain size is 160, 254, 260 and 455 nm, respectively.

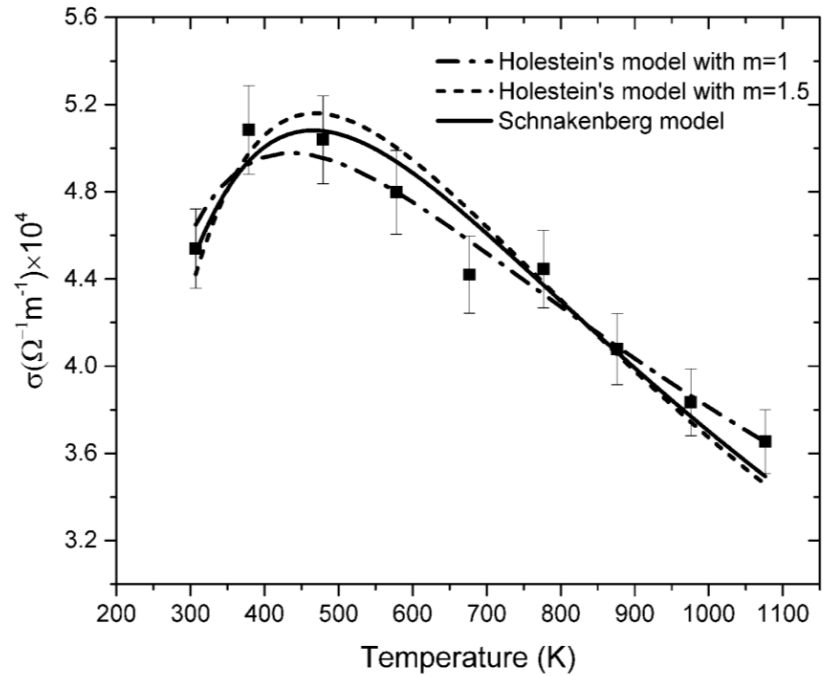

**Figure 4** Variation of electrical conductivity with temperature in $Ti_5O_9$ and fitted models.

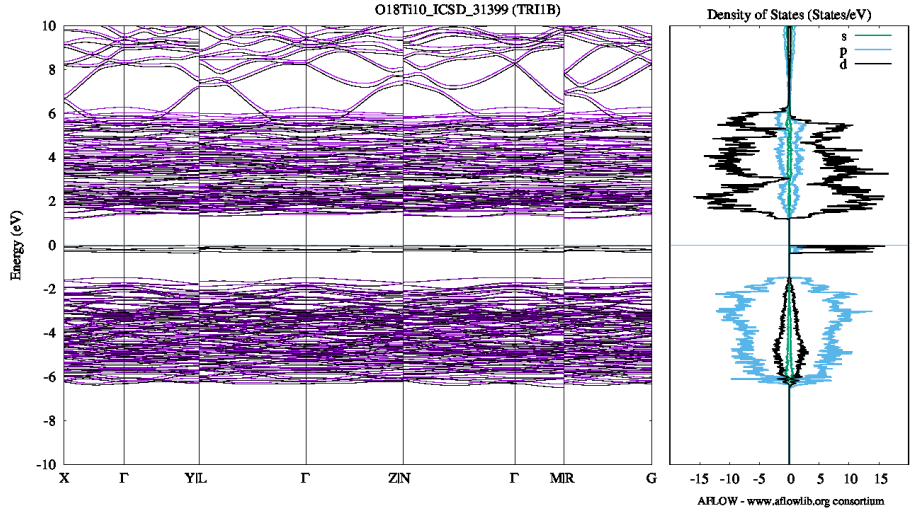

**Figure 5** Electronic band-structure of $Ti_5O_9$ as calculated by S. Curtarolo *et al*[5].

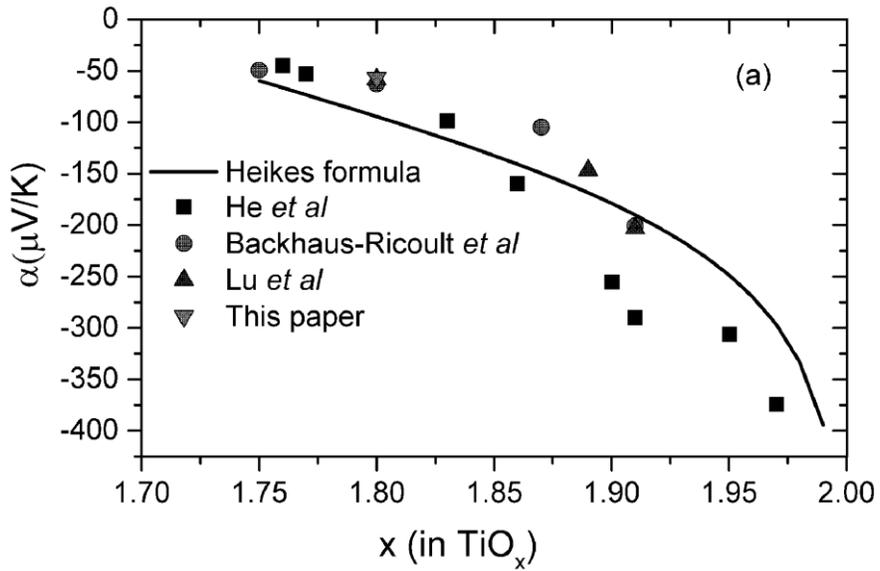

**Figure 6** Variation of the Seebeck coefficient with composition in $TiO_x$ compounds near ambient temperature. The solid line represents the Heikes-Chaikin-Beni model (Eq. 7b).

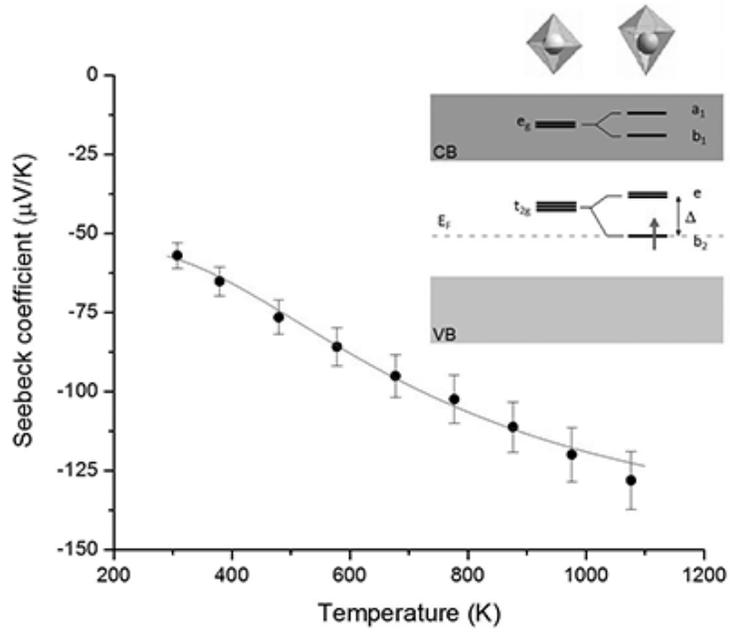

**Figure 6** Theoretical (solid curve) and experimental (dots) temperature dependency of the Seebeck coefficient in $Ti_5O_9$. The schematic in insert shows the energy levels arising from the effect of a $C_{4v}$ crystal-field on the 3d orbitals of $Ti^{3+}$. The lowest-lying $b_2$ state is singly-occupied and $\Delta$ denotes the energy separation between the singlet $b_2$ and the doublet e. The Fermi level is represented by the dashed line.

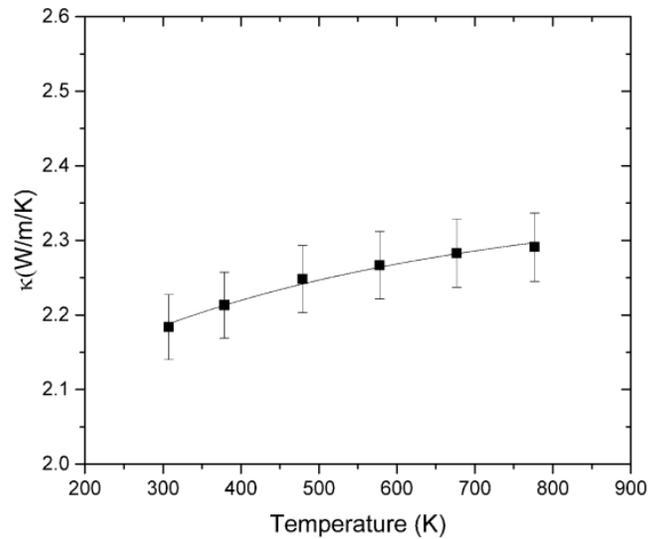

**Figure 7** Variation of the thermal conductivity of $Ti_5O_9$ with temperature.

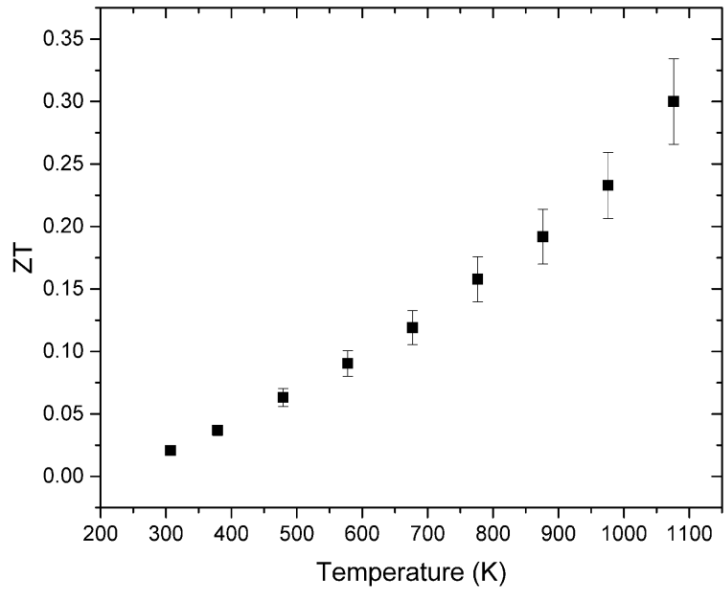

**Figure 8** Variation of ZT with temperature in $Ti_5O_9$.